\def\year{2022}\relax
\documentclass[letterpaper]{article} 
\usepackage{aaai22}  
\usepackage{times}  
\usepackage{helvet}  
\usepackage{courier}  
\usepackage[hyphens]{url}  
\usepackage{graphicx} 
\usepackage{natbib}  
\usepackage{caption} 
\DeclareCaptionStyle{ruled}{labelfont=normalfont,labelsep=colon,strut=off} 
\usepackage{algorithm}
\usepackage{algorithmic}

\usepackage{newfloat}
\usepackage{listings}
\floatstyle{ruled}
\newfloat{listing}{tb}{lst}{}
\floatname{listing}{Listing}
\usepackage{multirow}
\usepackage{amssymb}
\usepackage{amsmath}
\usepackage{subfigure}
\usepackage{array}
\usepackage{bm}

\title{Separated Contrastive Learning for Organ-at-Risk and Gross-Tumor-Volume Segmentation with Limited Annotation}%
\author{
    Jiacheng Wang\textsuperscript{\rm 1},
    Xiaomeng Li\textsuperscript{\rm 2},
    Yiming Han\textsuperscript{\rm 3},
    Jing Qin\textsuperscript{\rm 4},
    Liansheng Wang\textsuperscript{\rm 1,\thanks{Corresponding authors.}},
    Zhou Qichao\textsuperscript{\rm 5,\footnotemark[1]}
}
\affiliations {
    \textsuperscript{\rm 1} Department of Computer Science at School of Informatics, Xiamen University\\
    \textsuperscript{\rm 2} Department of Electronic and Computer Engineering, The Hong Kong University of Science and Technology\\
    \textsuperscript{\rm 3} Peking University 
    \textsuperscript{\rm 4} Center for Smart Health, School of Nursing, The Hong Kong Polytechnic University \\
    \textsuperscript{\rm 5} Manteia Technologies Co.,Ltd\\
    jiachengw@stu.xmu.edu.cn, eexmli@ust.hk, yiminghan\_emilia@foxmail.com, harry.qin@polyu.edu.hk, lswang@xmu.edu.cn, zhouqc@manteiatech.com
}

\begin{document}

\maketitle

\begin{abstract}
Automatic delineation of organ-at-risk (OAR) and gross-tumor-volume (GTV) is of great significance for radiotherapy planning.
However, it is a challenging task to learn powerful representations for accurate delineation under limited pixel (voxel)-wise annotations.
Contrastive learning at pixel-level can alleviate the dependency on annotations by learning dense representations from unlabeled data. 
Recent studies in this direction design various contrastive losses on the feature maps, to yield discriminative features for each pixel in the map.
However, pixels in the same map inevitably share semantics to be closer than they actually are, which may affect the discrimination of pixels in the same map and lead to the unfair comparison to pixels in other maps. 
To address these issues, we propose a separated region-level contrastive learning scheme, namely \emph{SepaReg}, the core of which is to separate each image into regions and encode each region separately.
Specifically, \emph{SepaReg} comprises two components: a structure-aware image separation (SIS) module and an intra- and inter-organ distillation (IID) module.
The SIS is proposed to operate on the image set to rebuild a region set under the guidance of structural information. 
The inter-organ representation will be learned from this set via typical contrastive losses cross regions.
On the other hand, the IID is proposed to tackle the quantity imbalance in the region set as tiny organs may produce fewer regions, by exploiting intra-organ representations.
We conducted extensive experiments to evaluate the proposed model on a public dataset and two private datasets.
The experimental results demonstrate the effectiveness of the proposed model, consistently achieving better performance than state-of-the-art approaches. 
Code is available at \url{https://github.com/jcwang123/Separate_CL}.
\end{abstract}


\begin{figure}[t]
    \centering
    \includegraphics[width=\linewidth]{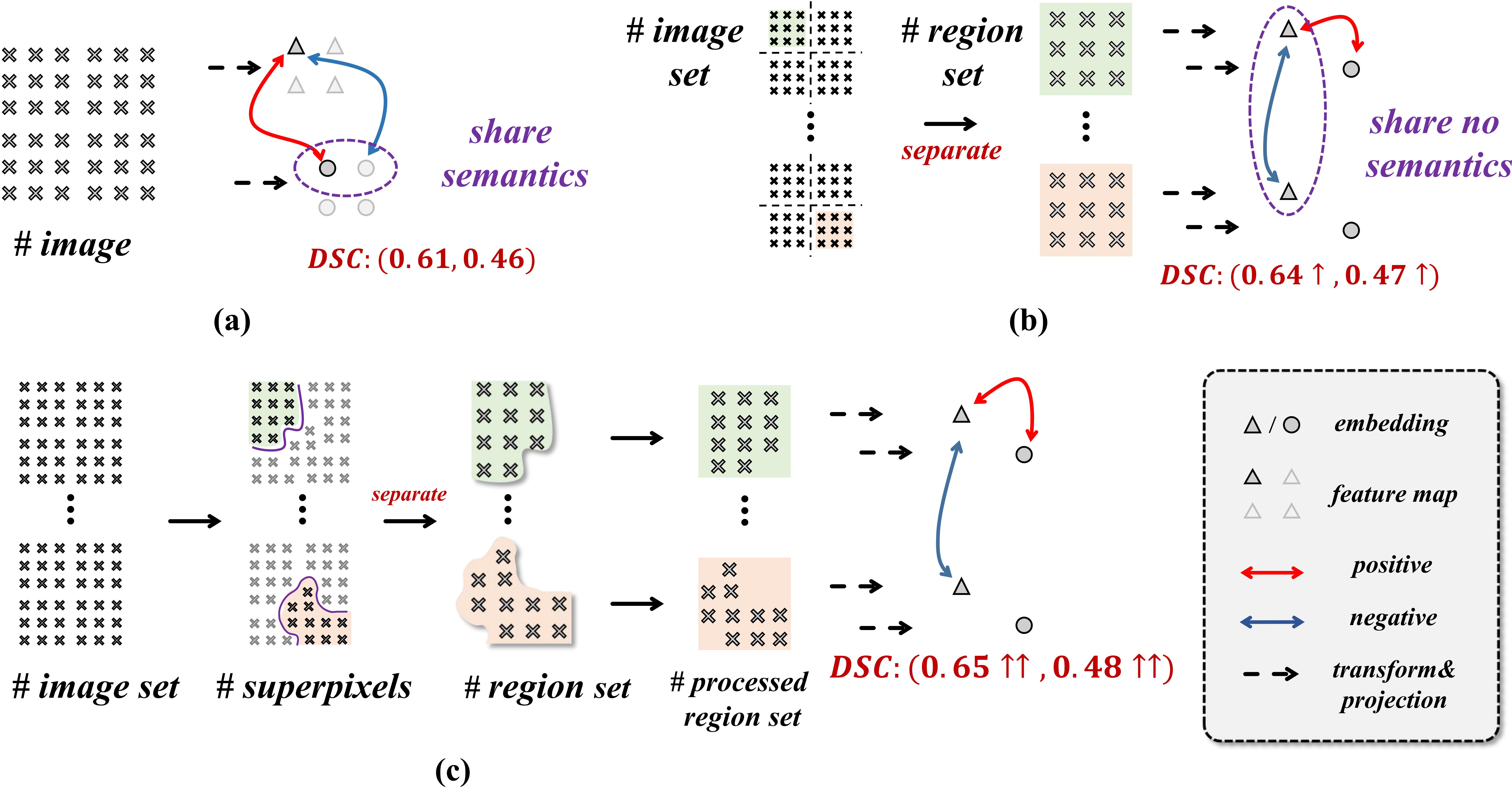}
    \caption{
    Motivation of building structure-aware image separation (SIS) module.
    (a) No separation, that forms comparison of pixels in the feature map. Pixels in the same map inevitably share semantics, which may affect the discrimination of pixels in the same map and lead to unfair comparison to pixels in other maps.
    (b) Regular separation, that regularly separates each image into square regions and encodes each region as a discriminative feature. Structure has been broken in the produced regions.
    (c) SIS, that separates each image into regions by superpixel-based division to learn their structural information. Note that the region set has been shuffled so that green region and orange region could be from different images. Evaluated DSC scores of two private datasets are shown in the figure.
    }
    \label{fig:learning_scheme}
\end{figure}
\section{Introduction}
\label{sec:introduction}
Accurate delineation of organ-at-risk (OAR) and gross-tumor-volume (GTV) in CT scans is a crucial step in radiotherapy treatment.
However, manual volume delineation is one of the most time-consuming and tedious tasks for clinicians.
Hence, automatic and accurate delineation tools are highly demanded in clinical practice. 
Recent years, researchers have proposed many deep models for medical image segmentation, such as U-Net~\cite{ronneberger2015u}, DenseUNet~\cite{li2018h}, HyperDenseNet~\cite{dolz2018hyperdense}, nnUNet~\cite{isensee2020nnu}. 
Successfully training these models usually requires a large number of pixel (voxel)-wise annotations.
However, in clinical practice, it is difficult, if not impossible, to acquire these annotations owing to the need of extensive professional expertise for labelling and the busy schedule of clinicians. 
In this regard, self-supervised learning, which attempts to learn discriminative features from a small set of labelled data and a large number of unlabeled data, becomes an important research direction in medical image segmentation. 

In recent years, contrastive learning~\cite{simclr,simclrv2,byol,simsam}, a learning strategy that first extracts features from the unlabeled dataset and then fine-tunes the network with a few labeled images, has dominated the field of self-supervised learning. 
The main idea of contrastive learning is to learn representations such that make similar samples stay close to each other while dissimilar ones far apart.
Nevertheless, when transferred into downstream tasks that require dense prediction, i.e., object segmentation, the representations learned by contrastive learning can only bring limited performance improvement, since it can only discriminate images rather than pixels.

Latest studies attempt to address this issue by learning discrimination of pixels in the feature maps~\cite{chaitanya2020contrastive,zhao2020contrastive,pixpro,chen2021bootstrap,alonso2021semi}. 
As shown in Figure~\ref{fig:learning_scheme}(a), after encoding an image into a feature map, each pixel in the map represents a square region in the image. 
Typical contrastive loss is then utilized to pull positive pixels together and push negative pixels far apart.
These methods differ in how to determine whether a pixel is positive or negative, by pseudo label~\cite{zhao2020contrastive,alonso2021semi} or spatial distance~\cite{chaitanya2020contrastive,pixpro,chen2021bootstrap}, yet all of them have neglected a pivotal problem that pixels in the feature map are sharing semantics, which is harmful to the pixel-level discrimination.
Specifically, this issue has negative effects on both intra-image and inter-image discrimination. 
%
First, the objective of intra-image discrimination is to distinguish an anchor pixel from other pixels in the same feature map; the sharing semantics, however, make them closer.
Second, the goal of inter-image discrimination is to find (dis-) similar pixels in other images, yet pixels from the same image are usually more similar than those from different images. 
The unfair comparison may destroy the exploration of pixel comparison cross images. 

To tackle this issue, we propose a novel separated region-level contrastive learning scheme for the challenging tasks of delineation of organ-at-risk (OAR) and gross-tumor-volume (GTV) in CT scans; we call the proposed model as \emph{SepaReg}.
It is composed of two components: a structure-aware image separation (SIS) module and an intra- and inter-organ distillation (IID) module.
\begin{itemize}
    \item \textbf{SIS} can be regarded as the core of the proposed \emph{SepaReg}, in which we propose to produce a brand-new region set and learn representation from the region set instead of the image set. Regions are produced with the guidance of structural information in order to keep complete anatomical structure to enhance the recognition of boundaries. The representations are learned by performing region comparison. Besides regions from the same image, each region is compared to more regions from other images due to the separation, effectively enhancing the regional diversity.

    \item \textbf{IID} is proposed to handle the quantity imbalance that larger organs will produce more regions, which may affect the representation learning of tiny organs~\cite{tian2021divide}. In IID, regions are clustered into different subsets according to their learned features; each subset represents a specific organ. The intra-organ representations are learned on each subset and distilled into one model in the end.
\end{itemize}

Extensive experiments are performed on three typical datasets, including one public dataset and two in-house datasets.
%
The two private datasets comprise OAR segmentation for lung cancer and GTV segmentation for nasopharyngeal cancer. 
We compare the proposed \emph{SepaReg} to several state-of-the-art image- and pixel-level contrastive learning schemes.
The transfer segmentation performance on a small labeled test set is used to assess the learned representation.
Experimental results demonstrate the effectiveness of the proposed model, consistently achieving better segmentation results than the state-of-the-art.

\section{Related Work}
\label{sec:related_work}
\subsection{Automatic Delineation in Radiation Therapy}
A handful of studies have been proposed to address CT-based automatic segmentation of OAR and GTV in the past few years.
Gradient-based image processing methods are adopted in the early years~\cite{geets2007gradient,day2009region,kerhet2010application}. They can give coarse segmentation results on OAR whose boundary is clear to identify.
While, it is still difficult to achieve satisfactory segmentation of GTV with ambiguous boundaries. 
Later, Deep Learning methods to perform OAR and GTV segmentation are rapidly developing and have caught lots of concentration. These models can easily obtain human-closed performance on most OARs of nasopharyngeal cancer~\cite{ibragimov2017segmentation}, lung cancer~\cite{zhu2019comparison}, and cervical cancer~\cite{liu2020segmentation}.
As for GTV segmentation, the most popular research direction is the fusion of different modalities, such as PET and CT~\cite{guo2019gross,jin2019accurate,jin2021deeptarget,2020Automated}.
However, models in these studies are trained under full supervision, accounting for precise annotations for each pixel (voxel). It is difficult to acquire these annotations thanks to the need of extensive professional expertise for labeling and the busy schedule of clinicians.
How to improve the segmentation performance given limited annotations, is still valuable to exploit till today. With this desirable consideration, we make a comprehensive effort to introduce a quite bleeding-edge technique, contrastive learning, in the field of label-efficient OAR and GTV segmentation.

\subsection{Representation Learning on CT Images}
The lack of sufficient expert-annotated data for model optimization, is one of the most general problems in the medical vision field due to the expensive cost in both time and experience.
Representation learning is a hot direction to solve this issue as it can help the model explore general representation from the unlabeled dataset, which can significantly improve the transfer performance on a relatively small labeled set. 
The representation could be learned by solving manually designed tasks, such as Jigsaw puzzle~\cite{zhuang2019self}, context restoration~\cite{taleb20203d}, orientation prediction~\cite{taleb20203d}, or their combination~\cite{zhou2019models,zhou2021models}. 
Most strategies work well on 3D networks but fail in 2D area~\cite{zhou2019models}, indicating the scant perception of semantic contexts in these representations.
The latest work has indicated the bright future of contrastive learning, which has dominated the field of representation learning in medical image segmentation~\cite{chaitanya2020contrastive}.
Despite its success in exploring global and local features, this work is still limited by the condition that all volumes should be aligned at first. 
Its setup for exploring pixel-level discrimination on feature maps also suffers from the sharing semantics.
By contrast, \emph{SepaReg} relies on no extra condition and avoids sharing semantics by its separated learning design.

\subsection{Contrastive Learning for Segmentation Tasks}
Contrastive learning is a successful and developing variant of representation learning, yet the investigation on segmentation tasks has not been fully studied.
Current work builds this scheme by forming pixel pairs on feature maps, according to pseudo label or spatial distance. 
The former is not suitable when given no class label during representation learning~\cite{zhao2020contrastive,alonso2021semi}, so that we mainly survey the latter.
The earliest work explores local feature comparison by determining dis-similar pixels in the feature map if they are far away in spatial dimension~\cite{chaitanya2020contrastive}.
Instead, PixPro ignores the dis-similar pixels and forms similar pixel pairs if they are closed~\cite{pixpro}.
Considering the spatial continuity of organs in sequential or volumetric images, pixels at the same location in neighbour slices are forced to stay close~\cite{zeng2021positional,chen2021bootstrap}. 
No matter what measurement they use, these studies adopt the same design that utilizes typical contrastive loss to discriminate each pixel in the feature map. 
However, they have ignored a vital problem that pixels in the same feature map inevitably share semantics to affect the pixel discrimination.
Our work aims to solve this issue through a separated region-level learning scheme, in which each region is encoded separately and shares no semantics. 

\subsection{Superpixel Segmentation}
Generally, superpixel is generated by clustering local pixels using low-level image properties such as color.
These methods are based on (1) graph-cut algorithms~\cite{felzenszwalb2004efficient,liu2011entropy} that treat image as un-directed graph and partition the graph based on edge-weights, or (2) clustering algorithms~\cite{achanta2010slic,neubert2014compact,li2015superpixel} such as $k$-means, that are initialized by seeding pixels and use color, spatial information to update the cluster centers. 

Based on superpixel, there is plenty of downstream application in medical imaging tasks~\cite{qin2018superpixel,spl_cls,ouyang2020self,li2020superpixel}.
It can be used for classification of hyperspectral images ~\cite{spl_cls}, organ segmentation in CT scans~\cite{qin2018superpixel,ouyang2020self}, label softening in brain MR images~\cite{li2020superpixel}, and so on.
Still, whether it is useful to introduce superpixel into the area of building dense contrastive learning has not been studied.
We propose to utilize the superpixel-based division to guide the model to learn structural information to enhance segmentation performance.
\begin{figure*}[t]
    \centering
    \includegraphics[width=\linewidth]{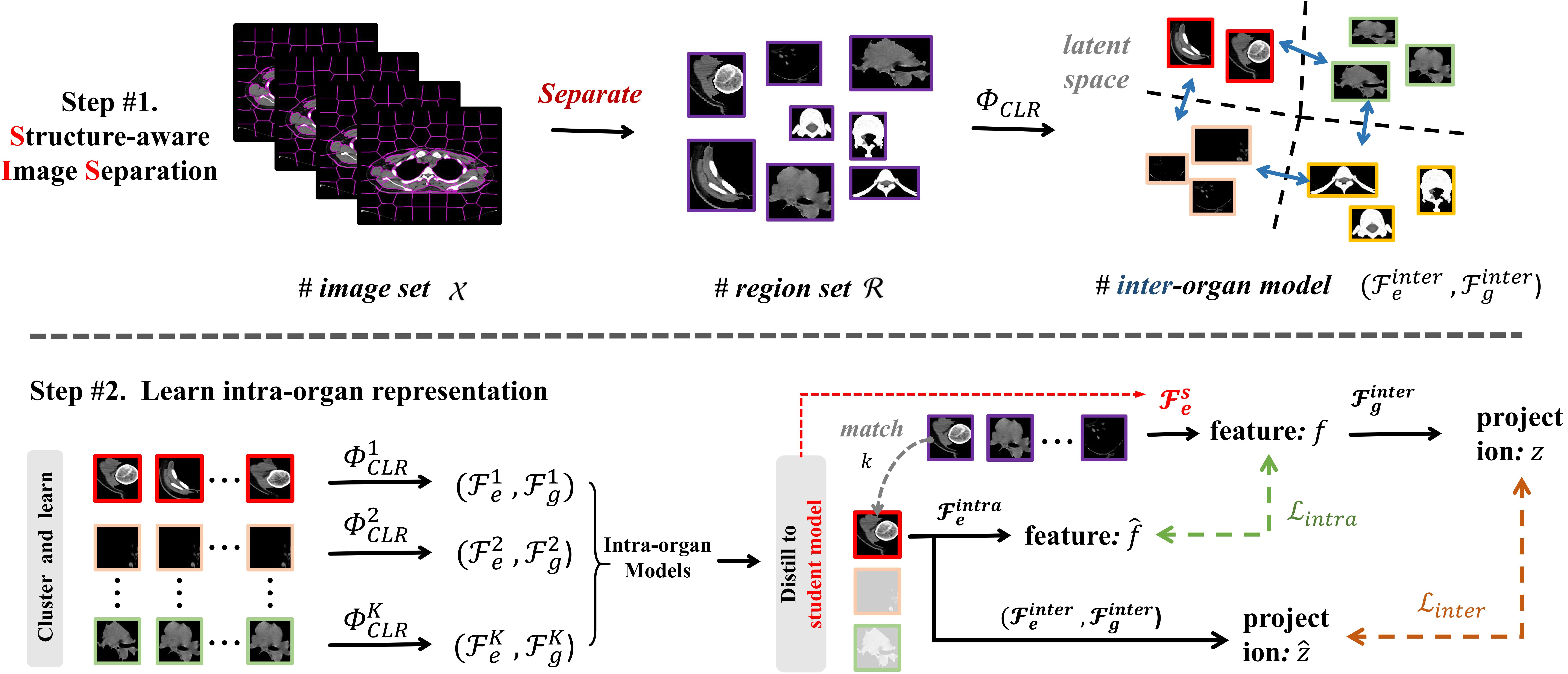}
    \caption{Framework of our separated region-level contrastive learning, \emph{SepaReg}, aimed at learning region-level representation in a separated manner. It comprises two major components: a structure-aware image separation (SIS) module and an intra- and inter-organ distillation (IID) module. SIS is proposed to solve the semantics sharing, by separating each image into several regions under the guidance of structural information, to form a brand-new region set and learn regions' representation from the set.
    IID is introduced to tackle the quantity imbalance in the region set since larger organ will produce more regions, by exploring intra-organ representations and distilling them into a student model.
    $\Phi_{CLR}$ denotes a standard contrastive learning operation, i.e., SimSiam. $\mathcal{F}_{e},\mathcal{F}_{g}$ denote the encoder and projector.
    }
    \label{fig:framework}
\end{figure*}
\section{Method}
\label{sec:method}
In this section, we will give a detailed description of our separated learning scheme. Overall framework has also been illustrated in Figure~\ref{fig:framework}. It contains two training steps, each of which will be described in an individual subsection.

\subsection{Structure-aware Image Separation}
\label{sec:sis}
Current pixel-level contrastive learning schemes encode images into feature maps and form pixel pairs in the maps. Differently, our approach starts by producing a set of regions from the image set; each region in the produced set will be embedded separately. We will describe the production in detail, followed by a brief introduction to the learning setup. 

Using superpixel to separate an image into regions for unsupervised segmentation is widely adopted in previous literature~\cite{qin2018superpixel,ouyang2020self}, yet not applied into the field of contrastive learning. 
Specifically, let $\mathcal{F}_s$ denote the superpixel operation. Given an image set $\mathcal{X}$, our target is to separate it into a set of regions $\mathcal{R}$:  $\mathcal{R} \gets \mathcal{F}_s(\mathcal{X})$.
Firstly, we separate each image $x$ in the image set into regions by SLIC method~\cite{achanta2010slic}. 
It begins with an initialization step where a certain number of cluster centers (set to 32 as default) are sampled on the regular grid spaced of $x$.
These centers are then moved to search locations with lowest gradient in a limited neighborhood, i.e., $3\times3$.
After the moving step, an update operation is performed to adjust the cluster centers.
The moving step and update operation will repeat until the new cluster centers move little, which is also similar to the $k$-means clustering algorithm.
After SLIC, each region produced from image $x$ will be padded with zero value to form a new square image for subsequent learning setup.
Till now, all produced regions from different images are unionized and shuffled to obtain the final region set $\mathcal{R}$.

To learn regions' representation, we perform a standard image-level contrastive learning operation on the region set.
This operation is denoted as $\Phi_{CLR}$, set to SimSiam~\cite{simsam} as default.
In formal, given a region $r$ that has been pre-processed with certain size ($128\times128$ as default), two augmentation views $v, v^{'}$ are created by different transformations $\mathcal{T},\mathcal{T}^{'}$.
The first augmented view $v$ is fed into an encoder $\mathcal{F}_{e}$, a standard ResNet-50 network~\cite{he2016deep} with removing its final two layers.
The extracted feature is then projected into the projection space via a projector $\mathcal{F}_{g}$, including two layers with a hidden dimension of 4096 and output size of 256.
Between the two layers, there is a sequence of Batch Normalization and Relu activation to avoid the collapsing solution~\cite{simsam}.
Similarly, another augmented view $v^{'}$ is fed into the encoder and projector to obtain its corresponding vector $z^{'}$.
An extensive predictor $\mathcal{F}_q$ with same architecture as projector is used to transfer $z^{'}$ into $z$ by regression.
The overall objective function is designed as 
\begin{equation}
\begin{gathered}
    \mathcal{L} = \frac{1}{2}(\mathcal{D}(\mathcal{F}_q(z), z^{'})+\mathcal{D}(z,\mathcal{F}_q(z^{'})))
\end{gathered}
\end{equation}
Here, $\mathcal{D}$ is used to measure the cosine similarity in the projection space, as
\begin{equation}
\label{eq:sim}
\mathcal{D}(\mathcal{F}_q(z), z^{'}) = -\frac{\mathcal{F}_q(z)}{||\mathcal{F}_q(z)||_2}\cdot\frac{z^{'}}{||z^{'}||_2},
\end{equation}
where $||\cdot||_2$ is $l_2$-norm. The error $\mathcal{L}$ is calculated for each region and the total loss is averaged on all regions in a mini-batch. 
As learned representation can coarsely discriminate different organs, we refer to it as the inter-organ model: $(\mathcal{F}_e^{inter},\mathcal{F}_g^{inter})$.

\subsection{Learn Intra-organ Representation}
\label{sec:d_regclr}
In this part, we extensively present a distillation-based module to solve the quantity imbalance by learning intra-organ representations. 
As the inter-organ model can tell semantic difference of organs, we propose to cluster the region set into several subsets according to what organs they are from.
For each organ, we will learn its specific representation by performing $\Phi_{CLR}$ on its corresponding subset so that tiny organs' representations could be learned well.
After that, we distill all intra-organ representations and the inter-organ representation into a student model. 

Formally, we cluster the region set into $K$ (empirically set to 5 as default) subsets according to their projection by $(\mathcal{F}_e^{inter},\mathcal{F}_g^{inter})$ based on $k$-means method, that are $\mathcal{R}^{1,2...K}$. 
For each subset, we will train a new encoder from scratch using the operator $\Phi_{CLR}$ to learn its intra-organ representation. In total, $K$ encoders will be trained at this stage, and we denote these intra-organ models as $\{(\mathcal{F}_e^{k},\mathcal{F}_g^{k})|k=1,2...K\}$.

The distillation module is trained under intra-organ and inter-organ regularization.
As shown in Figure~\ref{fig:framework}, it contains two parts:
(1) For intra-organ regularization, given a region $r$ from the region set $\mathcal{R}$, the organ it belongs to could be matched, as well as the intra-organ model corresponding to it. 
By feeding this region into the encoder of its intra-organ model, $\mathcal{F}_e^{intra}$, we could obtain the feature $\hat{f}$ that the student model should learn from. 
We minimize the $KL$-divergence error between two features:
\begin{equation}
    \mathcal{L}_{intra} = KL~(f~||~\hat{f}), 
\end{equation}
where $f$ is the feature encoded by the student model.
(2) For the inter-organ regularization, telling the discrimination between different organs is more important to learn. 
Therefore, we first feed the region into the inter-organ model to get its projection $\hat{z}$, then use the inter-organ projector to project $f$ into latent space and let the projection consistent with $\hat{z}$. 
The objective function is defined as 
\begin{equation}
    \mathcal{L}_{inter} = \mathcal{D}~(\mathcal{F}_{g}^{chair}(f)~,~\hat{z}),
\end{equation}
where $\mathcal{D}$ is the same measurement as Eq.\ref{eq:sim}.

In summary, $\mathcal{L}_{intra}$ is used to tell organ-specific feature representation and $\mathcal{L}_{inter}$ is used to learn the discriminative relation between different organs. 
We combine these two constraints to train our distillation network, that is, 
\begin{equation}
\label{eq:distill}
    \mathcal{L}_{distill} ~=~ \mathcal{L}_{intra}~+~\mathcal{L}_{inter}~.
\end{equation}

After distillation, we initialize a standard U-Net's encoder with the pretrained weight, followed by using Dice loss to optimize the parameters of the entire model. 
\begin{table*}[t]
\centering
\setlength\tabcolsep{3pt}
\small
\vspace{-2mm}
\begin{tabular}{cccc|ccc}
\hline
\multirow{2}{*}{Method}&\multicolumn{3}{c}{\textit{DSC}$\uparrow$}&\multicolumn{3}{c}{\textit{HD95}$\downarrow$}\\
\cline{2-7}
 & $|X_{tr}|=1$ & $|X_{tr}|=10$ & $|X_{tr}|=50$ & $|X_{tr}|=1$ & $|X_{tr}|=10$ & $|X_{tr}|=50$ \\ 
\hline
Random Init. &$61.26\pm0.55$ & $78.34\pm0.12$&$82.03\pm0.09 $&$6.49\pm6.94$ & $2.48\pm0.22$&$2.01\pm0.16$  \\
SimCLR~\cite{simclr} &$63.03\pm0.86$ &$80.88\pm0.13$&$83.61\pm0.08$
&$4.41\pm2.71$ &$2.07\pm0.10$&$1.80\pm0.18$ \\
BYOL~\cite{byol} &$61.66\pm0.61$ &$80.00\pm0.11$&$82.90\pm0.09$
&$5.12\pm2.42$ &$2.35\pm0.16$&$1.91\pm0.27$ \\
SimSiam~\cite{simsam} &$61.67\pm0.68$ &$79.68\pm0.11$&$83.16\pm0.09$
&$5.34\pm5.08$ &$2.22\pm0.20$&$1.92\pm0.36$ \\
GL~\cite{chaitanya2020contrastive}& $61.64\pm0.79$& $80.11\pm0.12$ &$82.60\pm0.09$
&$6.04\pm5.84$ &$2.28\pm0.15$&$1.84\pm0.21$ \\
PixPro~\cite{pixpro} & $65.34\pm0.51$& $79.65\pm0.12$ &$83.11\pm0.07$
&$5.13\pm3.58$ &$2.39\pm0.35$&$1.87\pm0.21$ \\
\hline
SepaReg &  \bm{$66.27\pm0.50$} &\bm{$81.59\pm0.13$} & \bm{$83.71\pm0.09$}
&\bm{$4.05\pm2.73$} & \bm{$1.90\pm0.13$} & \bm{$1.69\pm0.25$} \\
\hline
\end{tabular}
\caption{Comparison results on our private dataset, \textit{LungOAR}, with state-of-the-art contrastive learning methods, including image-level and pixel-level methods. "Random Init." means training from scratch. We show the DSC score (\%) and HD95 value (voxel) as well as the standard error in patient-wise.}
\label{tab:lungoar}
\end{table*}
%
%
%
%
\begin{table*}[t]
\centering
\renewcommand\arraystretch{0.9}
\setlength\tabcolsep{2pt}
\small
\vspace{-2mm}
\begin{tabular}{cccc|ccc}
\hline
\multirow{2}{*}{Method}&\multicolumn{3}{c}{\textit{DSC}$\uparrow$}&\multicolumn{3}{c}{\textit{HD95}$\downarrow$}\\
\cline{2-7}
 & $|X_{tr}|=1$ & $|X_{tr}|=10$ & $|X_{tr}|=50$ & $|X_{tr}|=1$ & $|X_{tr}|=10$ & $|X_{tr}|=50$ \\ 
\hline
Random Init. &$42.87\pm2.62$ & $55.28\pm3.19$&$61.05\pm2.61$ &$12.90\pm55.33$ & $8.90\pm39.58$&$11.11\pm234.00$  \\
SimCLR~\cite{simclr} &$47.41\pm2.98$ &$58.71\pm2.51$&$62.46\pm2.94$
&$12.96\pm45.53$ &$8.52\pm32.17$&$7.36\pm29.96$ \\
BYOL~\cite{byol} &$47.23\pm2.94$ &$59.01\pm2.54$&$63.62\pm2.81$
&$12.81\pm52.39$ &$8.23\pm28.98$&$7.32\pm31.35$ \\
SimSiam~\cite{simsam}&$46.06\pm2.94$ &$58.34\pm2.76$&$63.45\pm2.73$
&$14.81\pm56.71$ &$8.28\pm32.54$&$7.34\pm26.18$ \\
GL~\cite{chaitanya2020contrastive}& $48.30\pm2.71$& $59.06\pm2.91$ &$61.88\pm2.59$
&$13.70\pm85.04$&$8.25\pm32.44$&$7.25\pm31.66$ \\
PixPro~\cite{pixpro} &$48.52\pm2.82$ &$57.18\pm2.72$&$60.79\pm2.44$ 
&$12.92\pm53.45$ &$9.40\pm40.74$&$8.00\pm30.12$ \\
\hline
SepaReg & \bm{$50.30\pm2.12$} & \bm{$60.03\pm2.33$} & \bm{$63.98\pm2.06$}
&\bm{$12.31\pm44.06$} & \bm{$7.54\pm33.18$} & \bm{$6.77\pm30.27$} \\
\hline
\end{tabular}
\caption{Comparison results on our private dataset, \textit{NasoGTV}, with state-of-the-art contrastive learning methods, including image-level and pixel-level methods. "Random Init." means training from scratch. We show the DSC score (\%) and HD95 value (voxel) as well as the standard error in patient-wise.}
\label{tab:nasogtv}
\end{table*}
%
%
%
%
\begin{table*}[t]
\centering
\renewcommand\arraystretch{.9}
\small
\vspace{-2mm}
\setlength{\tabcolsep}{2pt}{
\begin{tabular}{c|c|cccccc}
\hline
\multirow{2}{*}{Method}&
\multirow{2}{*}{All OARs}&
\multicolumn{2}{c}{SMG} &
\multicolumn{2}{c}{Parotid}  &
\multirow{2}{*}{Brain Stem} & \multirow{2}{*}{Mandible} \\ 
&&Lt&Rt&Lt&Rt\\
\hline
Random Init. &$54.48\pm0.35$
&$38.61\pm3.99$&$25.03\pm2.27$
&$58.34\pm0.18$&$63.06\pm0.38$
&$64.45\pm2.15$&$77.38\pm0.19$\\
SimCLR & $57.10\pm0.22$
&$33.45\pm5.98$&$21.81\pm2.15$
&$62.30\pm0.34$&\bm{$68.23\pm0.25$}
&\bm{$75.11\pm0.28$}&\bm{$81.71\pm0.16$}\\
BYOL&$57.34\pm0.27$
&$40.21\pm6.49$&$25.32\pm0.86$
&$59.40\pm0.56$&$66.84\pm0.33$
&$73.63\pm0.61$&$78.65\pm0.32$\\
SimSiam &$54.83\pm0.14$ 
&$33.25\pm6.57$&$25.78\pm2.54$
&\bm{$65.79\pm0.23$}&$56.63\pm0.87$
&$68.25\pm1.38$&$79.27\pm0.23$\\
GL&$52.03\pm0.12$
&$44.06\pm2.54$&$6.01\pm0.35$
&$54.34\pm0.28$&$61.39\pm0.45$
&$65.05\pm0.79$&$81.30\pm0.20$\\
PixPro &$51.12\pm0.65$ 
&$37.78\pm5.04$&$18.40\pm1.57$
&$48.95\pm2.19$&$59.00\pm2.07$
&$62.26\pm1.65$&$80.33\pm0.26$\\
\hline
SepaReg &\bm{$58.43\pm0.10$} 
&\bm{$45.29\pm5.98$}&\bm{$27.99\pm1.55$}
&$58.42\pm0.35$&$67.65\pm0.16$
&$73.58\pm0.42$&$77.64\pm0.33$\\
\hline
\end{tabular}
}
\caption{Comparison results on a small public dataset, \textit{PDDCA}, with state-of-the-art contrastive learning methods, including image-level and pixel-level methods. We report the results when $|X_{tr}|=1$ as there is little available data. "Random Init." means training from scratch. We show the DSC score (\%) and standard error of each organ as well as the averaged value on all OARs in patient-wise.}
\label{tab:comparison_pubic}
\end{table*}

\section{Experiments}
\label{sec:experiments}

\subsection{Datasets}
We compare our method with several state-of-the-art contrastive learning methods on three CT datasets, including one public dataset and two in-house clinical datasets.

\noindent \textbf{PDDCA} is a public dataset consisting of 32 Head\&Neck CT scans with six OAR segmentation labels, i. e., submandibular gland (left and right), parotid (left and right), brain stem, and mandible~\cite{raudaschl2017evaluation}. 
However, the dataset is a bit small, lacks concrete and consistent annotations. Hence, we also collect two larger clinical datasets to validate the performance of our method. 

\noindent \textbf{LungOAR} is a clinical dataset, collected by Philips scanner in a local hospital. It contains 97 volumes and the target of this dataset is to segment the esophagus of lung cancer. For the annotation, one junior radiologist helps the first-round annotations, and one senior radiologist helps the second-round check to ensure accuracy. All data has been anonymized, and we have received approval from local hospitals for research purposes.

\noindent \textbf{NasoGTV} is a clinical dataset, consisting of 93 volumes. The images are collected by the CMS scanner. The target of this dataset is to segment the GTV of nasopharyngeal carcinoma. The same annotation process is performed by local radiologists and we have also received the approval.

\subsection{Experimental Setup and Comparison}
\label{sec:comparison}
\subsubsection{Pretrain stage} Following~\cite{chaitanya2020contrastive}, we split each dataset into a pre-training set $X_{pre}$ and a test set $X_{ts}$, where the volumetric images in $X_{pre}$ are used for pre-training, and those in $X_{ts}$ are only used to assess the segmentation performance. We randomly choose 77, 73, 22 volumes to form $X_{pre}$ for \textit{LungOAR} dataset, \textit{NasoGTV} dataset and \textit{PDDCA} dataset, respectively. The rest of volumes in each dataset will be used to form the test set.

\subsubsection{Finetune stage} As for the stage of fine-tuning, we choose a certain number of samples out of $X_{pre}$, consisting of both volumetric images and segmentation labels, to form the training set $X_{tr}$ and the validation set $X_{vl}$, where $|X_{vl}|=7$ in \textit{LungOAR}, $|X_{vl}|=3$ in \textit{NasoGTV}, and $|X_{vl}|=2$ in \textit{PDDCA}. We experiment with different sizes of $X_{tr}$ to assess the pre-trained representation. For instance, in \textit{LungOAR} and \textit{NasoGTV}, we build three experiments with $|X_{tr}|=1,10,50$. \textit{PDDCA} is so small that we only test the transfer performance with $|X_{tr}=1|$. 

We compare our method to several contrastive learning methods: (I) image-level methods: SimCLR~\cite{simclr}, BYOL~\cite{byol}, SimSiam~\cite{simsam}. (II) region-level methods: the scheme exploring global and local features, dubbed GL~\cite{chaitanya2020contrastive}, and PixPro~\cite{pixpro}.  
ResNet-50 and standard U-Net are used in all these experiments, and we implement these methods on three datasets by using the official code. 
We pre-train all the models for 100k iterations in total, and each mini-batch contains 32 images/regions. As for finetuning, we train the model for 200 epochs in all settings, save the best model of the validation set, and evaluate it on the test set in the end. Dice similarity coefficient (DSC) and 95\% Hausdorff Distance (HD95) are used to evaluate the segmentation performance at the patient level.  

\subsection{Comparison with the State-of-the-Art Methods}
Table~\ref{tab:lungoar} shows the results of our method on \textit{LungOAR} dataset. We can observe: (1) contrastive learning indeed brings improvements. Compared with "Random Init.", it is found that all methods have given better DSC score and HD95 value, and the improvement is larger with a smaller labeled set. (2) our method outperforms other methods with $1\%\sim2\%$ improvement on DSC score when $|X_{tr}|=1, 10$, and has also better result when $|X_{tr}|=50$. For example, when $|X_{tr}|=1$, our methods have improved the DSC score by $5.01\%$, which is significantly larger than the image-level method, i.e. $1.77\%$ by SimCLR, also obviously larger than region-level method, i.e., $4.08\%$ by PixPro. (3) it is also noteworthy that PixPro can outperform other image-level contrastive learning methods when $|X_{tr}|=1$, while show inferior results when enlarging the training set. It indicates that the representation learned by PixPro is not suitable for the downstream segmentation task. In contrast, our method brings larger improvement even when $|X_{tr}|=50$.

As shown in Table~\ref{tab:nasogtv}, the results on \textit{NasoGTV} keep consistent with those from Table~\ref{tab:lungoar}, validating the effectiveness of our method to extract self-learned features for the segmentation task. 
Since GTV segmentation is more challenging than OAR segmentation, we note that our method achieves outstanding improvements.


Table~\ref{tab:comparison_pubic} summarizes the results on PDDCA dataset. We can observe the best average score of all OARs achieved by our method.
The results further indicate the robustness of our method. It is noteworthy that GL and PixPro yield worse segmentation results compared to the random initialization. This is because organs in this dataset have no square shape, i.e., parotid that is spindly, leading to mistakes in obtaining the region pairs. 
Instead, our proposed separated learning scheme can take advantage of structural information and thus enable the network to learn shape-adaptive knowledge.

\begin{table}[t]
\centering
\renewcommand\arraystretch{1}
\vspace{-2mm}
\begin{tabular}{ccc}
\hline
  & \textit{LungOAR} & \textit{NasoGTV} \\
\hline
 \textit{no separation}  & $60.63\pm0.97$ & $46.22\pm2.58$\\
 \textit{regular separation} & $63.78\pm 0.69$ & $47.14\pm2.91$ \\
\hline
\textit{SIS} & $65.32\pm 0.48$ & $48.41\pm2.79$ \\
\hline
\end{tabular}
\caption{Comparison of different separation strategies: no separation, regular separation, and structure-aware separation (SIS). We report the DSC score (\%) and standard error in patient-wise.}
\label{tab:sepa}
\end{table}

\begin{table}[t]
\centering
\renewcommand\arraystretch{1}
\vspace{-2mm}
\begin{tabular}{m{14mm}<{\centering}m{14mm}<{\centering}cc}
\hline
\textit{SIS} & \textit{IID} & \textit{LungOAR} & \textit{NasoGTV} \\
\hline
 & & $61.26\pm0.55$& $42.87\pm2.62$ \\
\checkmark  & &$65.32\pm0.48$& $48.41\pm2.79$ \\
\checkmark  & \checkmark& $66.27\pm0.50$& $50.30\pm2.12$ \\
\hline
\end{tabular}
\caption{Ablation study of SIS and IID. The first row with no modules denotes training with random initialization. The combination of SIS and IID, is our separated region-level contrastive learning scheme, SepaReg. We report the DSC value (\%) and standard error in patient-wise.}
\label{tab:independence}
\end{table}

\begin{table}[t]
\centering
\renewcommand\arraystretch{1}
\vspace{-2mm}
\begin{tabular}{m{14mm}<{\centering}m{14mm}<{\centering}cc}
\hline
 $\mathcal{L}_{intra}$ & $\mathcal{L}_{inter}$ & \textit{LungOAR} & \textit{NasoGTV} \\
\hline
 & & $65.32\pm0.48$ & $48.41\pm2.79$\\
\checkmark & & $65.80\pm0.65$ & $48.79\pm2.82$ \\
\checkmark & \checkmark & $66.27\pm0.50$ & $50.30\pm2.12$ \\
\hline
\end{tabular}
\caption{Analytical study of two objectives in our distillation network. Basic scheme without any objectives is the separate learning scheme using superpixel method.}
\label{tab:distillation}
\end{table}
  
\subsection{Ablation Study}
In this part, we analyze the effectiveness of each module in our method.
For simplification, we only present the DSC score when $|X_{tr}|=1$. 

\subsubsection{How important the structure-aware separation is?}
As shown in Table~\ref{tab:sepa}, we first compare with three learning schemes to verify the effectiveness of superpixel-based separable feature extraction.  
They have also been visualized in Figure~\ref{fig:learning_scheme}, and the result has been shown in Table~\ref{tab:sepa}. \emph{\textbf{No separation}} refers to form positive region pairs at the same location of two feature maps. It can be regarded as PixPro with removing its PPM module. \emph{\textbf{Regular separation}} refer to separate each image into regular grid regions, and \emph{\textbf{SIS}} is the structure-aware image separation. We can see that the separation can improve transfer segmentation performance, demonstrating the effectiveness of separating the feature extraction. SIS extensively increases the DSC scores, indicating that structural-aware separation can extract structural information for better segmentation. 

\subsubsection{How do two major components affect?}
\emph{SepaReg} is composed of two major parts, and we make an analytical study about their influence in Table~\ref{tab:independence}. Firstly, compared to the model trained from scratch, it is observed that the SIS has already yielded good segmentation improvement. The DSC value has been improved by $4.06\%$ and $5.54\%$, indicating the powerful ability in learning region-level representation. Secondly, after learning intra-organ representations, the transfer performance reaches the best, since IID can break the quantity imbalance between regions of different organs.

\subsubsection{Design of intra- and inter-organ distillation.}
We conduct an ablation experiment about these two objectives of IID, $\mathcal{L}_{intra}$ and $\mathcal{L}_{inter}$, in Table~\ref{tab:distillation}. It could be seen that constraints from the intra-organ models can significantly improve the transfer performance and constraint from inter-organ model can further improve it, demonstrating the complementary advantage of both objectives.

\subsection{Other Analysis}
\label{sec:analysis}
\subsubsection{How does training iteration matter?}
\begin{figure}[t]
    \centering
    \includegraphics[width=\linewidth]{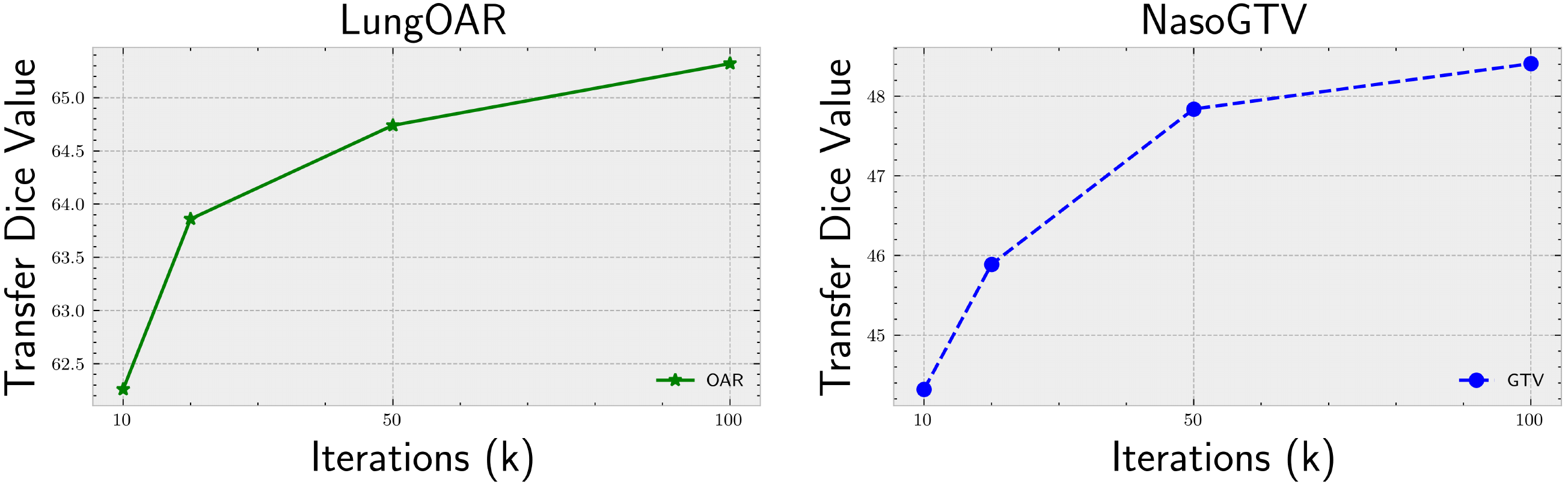}
    \caption{The transfer performance of models pre-trained at different amount of iterations on OAR and GTV datasets (left: \textit{LungOAR}; right: \textit{NasoGTV}).}
    \label{fig:iter}
\end{figure}
Dividing images into regions can help the model learn to discriminate different regions. Still, it needs more iterations to update the parameters, as one mini-batch of regions contains much less information than that of images. We evaluate the transfer segmentation performance of pretrained weights at different iterations, i.e. 10k, 50k and 100k. According to the result in Figure~\ref{fig:iter}, it is observed that when trained at a small number of iterations, SIS performs not so good. While meeting more samples and trained at larger iterations, the performance grows to the best.

\begin{figure}[h]
    \centering
    \includegraphics[width=1\linewidth]{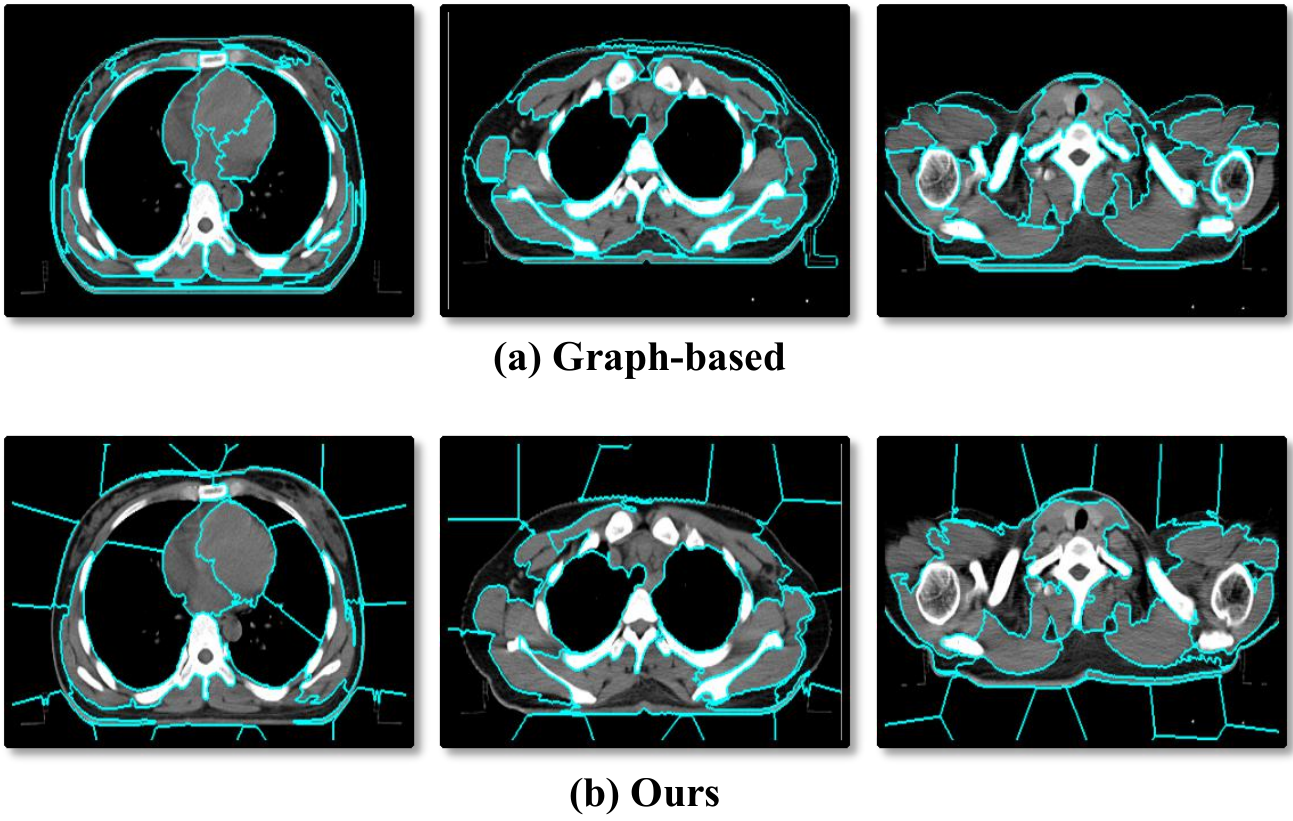}
    \caption{Regions in lung CT scans produced by (a) gragh-based superpixel method, and (b) our method. It's found that the gragh-based method can produce regions that best fit the boundary but have large variety in shape and size, not suitable for subsequent learning scheme.
    Regions produced by our method have more regular shape and can also provide complete structural information to discriminate different organs, i.e., lung and heart.
    %
    }
    \label{fig:super}
\end{figure}
\subsubsection{Different superpixel methods.} 
SIS applies a cluster-based superpixel method to guide the image separation. However, some work argues that using the graph-based method can represent the structural information better~\cite{ouyang2020self}. We give several typical results generated by two types of superpixel methods in Figure~\ref{fig:super}. It is found that the graph-based methods can yield regions that fit the boundary better than our produced regions. Despite it, regions created by the graph-based method have more irregular shapes and sizes at the same time, not suitable for subsequent learning schemes.

\begin{table}[t]
\centering
\renewcommand\arraystretch{1}
\begin{tabular}{m{14mm}<{\centering}m{14mm}<{\centering}cc}
\hline
 SepaReg & M\&T & \textit{LungOAR} & \textit{NasoGTV} \\
\hline
\checkmark &  & $66.27\pm0.50$ & $50.30\pm2.12$\\
 & \checkmark & $62.92\pm 0.48$ & $45.19\pm2.58$ \\
\checkmark & \checkmark & $67.11\pm0.30$ & $50.80\pm2.80$ \\
\hline
\end{tabular}
\caption{Extensive experiment about the comparison of SepaReg and Mean Teacher (M\&T). Here, we report the DSC score (\%) and standar error in patient-wise.}
\label{tab:semi}
\end{table}

\subsubsection{Combination with semi-supervised method.}
Contrastive learning aims to provide suitable initialized parameters for better transfer performance on a small amount of labeled data. With the same goal, semi-supervised method, i.e., Mean Teacher, is also helpful. In this part, we make an extensive experiment to verify two points: (i) \emph{SepaReg} can give better segmentation performance on limited labeled data, (ii) \emph{SepaReg} and Mean Teacher can provide complementary information to boost the segmentation performance further. Specifically, we construct this experiment on both datasets with $|X_{tr}|=1$. The same unlabeled dataset is used for Mean Teacher (M\&T) training ($X_{pre}$). We combine \emph{SepaReg} and M\&T by initializing the "Teacher-Student" model in M\&T with parameters trained by \emph{SepaReg}. 
It is found that (1) on both datasets, \emph{SepaReg} outperforms Mean Teacher significantly, i.e. $3.35\%$ and $5.11\%$ improvements on DSC sore, and (2) the combination can further improve the DSC scores, indicating the valuable potential of their complementary information. 
\section{Conclusion}
In this paper, we devise a separated region-level contrastive learning scheme, named SepaReg, to solve the problem of sharing semantics in latest pixel-level contrastive learning schemes. SepaReg comprises two components: structure-aware image separation (SIS) and intra- and inter-organ distillation module (IID). SIS proposes to produce a brand-new region set by separating each image into regions under the guidance of structural information, and learn their representations by forming region comparison. IID is proposed to boost the representation learning of tiny organs, since tiny organs produce few regions in the set, by exploring inra-organ representations. SepaReg is evaluated on one public dataset and two clinical datasets, achieving the best DSC score and HD95 value on all test sets, compared to several image- and pixel-level contrastive learning methods.
\section*{Acknowledgements} 
This work was supported by Ministry of Science and Technology of the People's Republic of China (2021ZD0201900) (2021ZD0201903).

\bibliography{aaai22}

\begin{thebibliography}{40}
\providecommand{\natexlab}[1]{#1}

\bibitem[{Achanta et~al.(2010)Achanta, Shaji, Smith, Lucchi, Fua, and
  S{\"u}sstrunk}]{achanta2010slic}
Achanta, R.; Shaji, A.; Smith, K.; Lucchi, A.; Fua, P.; and S{\"u}sstrunk, S.
  2010.
\newblock Slic superpixels.
\newblock Technical report.

\bibitem[{Alonso et~al.(2021)Alonso, Sabater, Ferstl, Montesano, and
  Murillo}]{alonso2021semi}
Alonso, I.; Sabater, A.; Ferstl, D.; Montesano, L.; and Murillo, A.~C. 2021.
\newblock Semi-Supervised Semantic Segmentation with Pixel-Level Contrastive
  Learning from a Class-wise Memory Bank.
\newblock \emph{arXiv preprint arXiv:2104.13415}.

\bibitem[{Chaitanya et~al.(2020)Chaitanya, Erdil, Karani, and
  Konukoglu}]{chaitanya2020contrastive}
Chaitanya, K.; Erdil, E.; Karani, N.; and Konukoglu, E. 2020.
\newblock Contrastive learning of global and local features for medical image
  segmentation with limited annotations.
\newblock \emph{Advances in Neural Information Processing Systems}, 33.

\bibitem[{Chen et~al.(2020{\natexlab{a}})Chen, Kornblith, Norouzi, and
  Hinton}]{simclr}
Chen, T.; Kornblith, S.; Norouzi, M.; and Hinton, G. 2020{\natexlab{a}}.
\newblock A simple framework for contrastive learning of visual
  representations.
\newblock In \emph{International conference on machine learning}, 1597--1607.
  PMLR.

\bibitem[{Chen et~al.(2020{\natexlab{b}})Chen, Kornblith, Swersky, Norouzi, and
  Hinton}]{simclrv2}
Chen, T.; Kornblith, S.; Swersky, K.; Norouzi, M.; and Hinton, G.
  2020{\natexlab{b}}.
\newblock Big self-supervised models are strong semi-supervised learners.
\newblock \emph{arXiv preprint arXiv:2006.10029}.

\bibitem[{Chen and He(2020)}]{simsam}
Chen, X.; and He, K. 2020.
\newblock Exploring Simple Siamese Representation Learning.
\newblock \emph{arXiv preprint arXiv:2011.10566}.

\bibitem[{Chen et~al.(2021)Chen, Zhuo, Wang, Xue, and Ni}]{chen2021bootstrap}
Chen, Z.; Zhuo, W.; Wang, T.; Xue, W.; and Ni, D. 2021.
\newblock Bootstrap Representation Learning for Segmentation on Medical Volumes
  and Sequences.
\newblock \emph{arXiv preprint arXiv:2106.12153}.

\bibitem[{Day et~al.(2009)Day, Betler, Parda, Reitz, Kirichenko, Mohammadi, and
  Miften}]{day2009region}
Day, E.; Betler, J.; Parda, D.; Reitz, B.; Kirichenko, A.; Mohammadi, S.; and
  Miften, M. 2009.
\newblock A region growing method for tumor volume segmentation on PET images
  for rectal and anal cancer patients.
\newblock \emph{Medical physics}, 36(10): 4349--4358.

\bibitem[{Dolz et~al.(2018)Dolz, Gopinath, Yuan, Lombaert, Desrosiers, and
  Ayed}]{dolz2018hyperdense}
Dolz, J.; Gopinath, K.; Yuan, J.; Lombaert, H.; Desrosiers, C.; and Ayed, I.~B.
  2018.
\newblock HyperDense-Net: a hyper-densely connected CNN for multi-modal image
  segmentation.
\newblock \emph{IEEE transactions on medical imaging}, 38(5): 1116--1126.

\bibitem[{Felzenszwalb and Huttenlocher(2004)}]{felzenszwalb2004efficient}
Felzenszwalb, P.~F.; and Huttenlocher, D.~P. 2004.
\newblock Efficient graph-based image segmentation.
\newblock \emph{International journal of computer vision}, 59(2): 167--181.

\bibitem[{Geets et~al.(2007)Geets, Lee, Bol, Lonneux, and
  Gr{\'e}goire}]{geets2007gradient}
Geets, X.; Lee, J.~A.; Bol, A.; Lonneux, M.; and Gr{\'e}goire, V. 2007.
\newblock A gradient-based method for segmenting FDG-PET images: methodology
  and validation.
\newblock \emph{European journal of nuclear medicine and molecular imaging},
  34(9): 1427--1438.

\bibitem[{Grill et~al.(2020)Grill, Strub, Altch{\'e}, Tallec, Richemond,
  Buchatskaya, Doersch, Pires, Guo, Azar et~al.}]{byol}
Grill, J.-B.; Strub, F.; Altch{\'e}, F.; Tallec, C.; Richemond, P.~H.;
  Buchatskaya, E.; Doersch, C.; Pires, B.~A.; Guo, Z.~D.; Azar, M.~G.; et~al.
  2020.
\newblock Bootstrap your own latent: A new approach to self-supervised
  learning.
\newblock \emph{arXiv preprint arXiv:2006.07733}.

\bibitem[{Guo et~al.(2019)Guo, Guo, Gong, Li et~al.}]{guo2019gross}
Guo, Z.; Guo, N.; Gong, K.; Li, Q.; et~al. 2019.
\newblock Gross tumor volume segmentation for head and neck cancer radiotherapy
  using deep dense multi-modality network.
\newblock \emph{Physics in Medicine \& Biology}, 64(20): 205015.

\bibitem[{He et~al.(2016)He, Zhang, Ren, and Sun}]{he2016deep}
He, K.; Zhang, X.; Ren, S.; and Sun, J. 2016.
\newblock Deep residual learning for image recognition.
\newblock In \emph{Proceedings of the IEEE conference on computer vision and
  pattern recognition}, 770--778.

\bibitem[{Ibragimov and Xing(2017)}]{ibragimov2017segmentation}
Ibragimov, B.; and Xing, L. 2017.
\newblock Segmentation of organs-at-risks in head and neck CT images using
  convolutional neural networks.
\newblock \emph{Medical physics}, 44(2): 547--557.

\bibitem[{Isensee et~al.(2020)Isensee, Jaeger, Kohl, Petersen, and
  Maier-Hein}]{isensee2020nnu}
Isensee, F.; Jaeger, P.~F.; Kohl, S.~A.; Petersen, J.; and Maier-Hein, K.~H.
  2020.
\newblock nnU-Net: a self-configuring method for deep learning-based biomedical
  image segmentation.
\newblock \emph{Nature Methods}, 1--9.

\bibitem[{Jia et~al.(2020)Jia, Deng, Xu, Zhou, and Jia}]{spl_cls}
Jia, S.; Deng, X.; Xu, M.; Zhou, J.; and Jia, X. 2020.
\newblock Superpixel-Level Weighted Label Propagation for Hyperspectral Image
  Classification.
\newblock \emph{IEEE Transactions on Geoscience and Remote Sensing}, 58(7):
  5077--5091.

\bibitem[{Jin et~al.(2019)Jin, Guo, Ho, Harrison, Xiao, Tseng, and
  Lu}]{jin2019accurate}
Jin, D.; Guo, D.; Ho, T.-Y.; Harrison, A.~P.; Xiao, J.; Tseng, C.-K.; and Lu,
  L. 2019.
\newblock Accurate esophageal gross tumor volume segmentation in pet/ct using
  two-stream chained 3d deep network fusion.
\newblock In \emph{International Conference on Medical Image Computing and
  Computer-Assisted Intervention}, 182--191. Springer.

\bibitem[{Jin et~al.(2021)Jin, Guo, Ho, Harrison, Xiao, Tseng, and
  Lu}]{jin2021deeptarget}
Jin, D.; Guo, D.; Ho, T.-Y.; Harrison, A.~P.; Xiao, J.; Tseng, C.-K.; and Lu,
  L. 2021.
\newblock DeepTarget: Gross tumor and clinical target volume segmentation in
  esophageal cancer radiotherapy.
\newblock \emph{Medical Image Analysis}, 68: 101909.

\bibitem[{Kerhet et~al.(2010)Kerhet, Small, Quon, Riauka, Schrader, Greiner,
  Yee, McEwan, and Roa}]{kerhet2010application}
Kerhet, A.; Small, C.; Quon, H.; Riauka, T.; Schrader, L.; Greiner, R.; Yee,
  D.; McEwan, A.; and Roa, W. 2010.
\newblock Application of machine learning methodology for PET-based definition
  of lung cancer.
\newblock \emph{Current oncology}, 17(1): 41.

\bibitem[{Li et~al.(2020)Li, Wei, Cao, Ma, Wang, and Zheng}]{li2020superpixel}
Li, H.; Wei, D.; Cao, S.; Ma, K.; Wang, L.; and Zheng, Y. 2020.
\newblock Superpixel-Guided Label Softening for Medical Image Segmentation.
\newblock In \emph{International Conference on Medical Image Computing and
  Computer-Assisted Intervention}, 227--237. Springer.

\bibitem[{Li et~al.(2018)Li, Chen, Qi, Dou, Fu, and Heng}]{li2018h}
Li, X.; Chen, H.; Qi, X.; Dou, Q.; Fu, C.-W.; and Heng, P.-A. 2018.
\newblock H-DenseUNet: hybrid densely connected UNet for liver and tumor
  segmentation from CT volumes.
\newblock \emph{IEEE transactions on medical imaging}, 37(12): 2663--2674.

\bibitem[{Li and Chen(2015)}]{li2015superpixel}
Li, Z.; and Chen, J. 2015.
\newblock Superpixel segmentation using linear spectral clustering.
\newblock In \emph{Proceedings of the IEEE Conference on Computer Vision and
  Pattern Recognition}, 1356--1363.

\bibitem[{Liu et~al.(2011)Liu, Tuzel, Ramalingam, and
  Chellappa}]{liu2011entropy}
Liu, M.-Y.; Tuzel, O.; Ramalingam, S.; and Chellappa, R. 2011.
\newblock Entropy rate superpixel segmentation.
\newblock In \emph{CVPR 2011}, 2097--2104. IEEE.

\bibitem[{Liu et~al.(2020)Liu, Liu, Xiao, Wang, Miao, Sun, and
  Zhang}]{liu2020segmentation}
Liu, Z.; Liu, X.; Xiao, B.; Wang, S.; Miao, Z.; Sun, Y.; and Zhang, F. 2020.
\newblock Segmentation of organs-at-risk in cervical cancer CT images with a
  convolutional neural network.
\newblock \emph{Physica Medica}, 69: 184--191.

\bibitem[{Neubert and Protzel(2014)}]{neubert2014compact}
Neubert, P.; and Protzel, P. 2014.
\newblock Compact watershed and preemptive slic: On improving trade-offs of
  superpixel segmentation algorithms.
\newblock In \emph{2014 22nd international conference on pattern recognition},
  996--1001. IEEE.

\bibitem[{Ouyang et~al.(2020)Ouyang, Biffi, Chen, Kart, Qiu, and
  Rueckert}]{ouyang2020self}
Ouyang, C.; Biffi, C.; Chen, C.; Kart, T.; Qiu, H.; and Rueckert, D. 2020.
\newblock Self-supervision with Superpixels: Training Few-Shot Medical Image
  Segmentation Without Annotation.
\newblock In \emph{European Conference on Computer Vision}, 762--780. Springer.

\bibitem[{Qin et~al.(2018)Qin, Wu, Han, Yuan, Zhao, Ibragimov, Gu, and
  Xing}]{qin2018superpixel}
Qin, W.; Wu, J.; Han, F.; Yuan, Y.; Zhao, W.; Ibragimov, B.; Gu, J.; and Xing,
  L. 2018.
\newblock Superpixel-based and boundary-sensitive convolutional neural network
  for automated liver segmentation.
\newblock \emph{Physics in Medicine \& Biology}, 63(9): 095017.

\bibitem[{Raudaschl et~al.(2017)Raudaschl, Zaffino, Sharp, Spadea, Chen,
  Dawant, Albrecht, Gass, Langguth, L{\"u}thi et~al.}]{raudaschl2017evaluation}
Raudaschl, P.~F.; Zaffino, P.; Sharp, G.~C.; Spadea, M.~F.; Chen, A.; Dawant,
  B.~M.; Albrecht, T.; Gass, T.; Langguth, C.; L{\"u}thi, M.; et~al. 2017.
\newblock Evaluation of segmentation methods on head and neck CT:
  auto-segmentation challenge 2015.
\newblock \emph{Medical physics}, 44(5): 2020--2036.

\bibitem[{Ronneberger, Fischer, and Brox(2015)}]{ronneberger2015u}
Ronneberger, O.; Fischer, P.; and Brox, T. 2015.
\newblock U-net: Convolutional networks for biomedical image segmentation.
\newblock In \emph{International Conference on Medical image computing and
  computer-assisted intervention}, 234--241. Springer.

\bibitem[{Taleb et~al.(2020)Taleb, Loetzsch, Danz, Severin, Gaertner, Bergner,
  and Lippert}]{taleb20203d}
Taleb, A.; Loetzsch, W.; Danz, N.; Severin, J.; Gaertner, T.; Bergner, B.; and
  Lippert, C. 2020.
\newblock 3D Self-Supervised Methods for Medical Imaging.
\newblock \emph{arXiv preprint arXiv:2006.03829}.

\bibitem[{Tian, Henaff, and van~den Oord(2021)}]{tian2021divide}
Tian, Y.; Henaff, O.~J.; and van~den Oord, A. 2021.
\newblock Divide and Contrast: Self-supervised Learning from Uncurated Data.
\newblock arXiv:2105.08054.

\bibitem[{Wang et~al.(2020)Wang, Yang, Zhang, Zhu, and Dai}]{2020Automated}
Wang, X.; Yang, G.; Zhang, Y.; Zhu, L.; and Dai, Z. 2020.
\newblock Automated delineation of nasopharynx gross tumor volume for
  nasopharyngeal carcinoma by plain CT combining contrast-enhanced CT using
  deep learning.
\newblock \emph{Journal of Radiation Research and Applied Sciences}, 13(1):
  568--577.

\bibitem[{Xie et~al.(2020)Xie, Lin, Zhang, Cao, Lin, and Hu}]{pixpro}
Xie, Z.; Lin, Y.; Zhang, Z.; Cao, Y.; Lin, S.; and Hu, H. 2020.
\newblock Propagate Yourself: Exploring Pixel-Level Consistency for
  Unsupervised Visual Representation Learning.
\newblock \emph{arXiv preprint arXiv:2011.10043}.

\bibitem[{Zeng et~al.(2021)Zeng, Wu, Hu, Xu, Yuan, Huang, Zhuang, Hu, and
  Shi}]{zeng2021positional}
Zeng, D.; Wu, Y.; Hu, X.; Xu, X.; Yuan, H.; Huang, M.; Zhuang, J.; Hu, J.; and
  Shi, Y. 2021.
\newblock Positional Contrastive Learning for Volumetric Medical Image
  Segmentation.
\newblock \emph{arXiv preprint arXiv:2106.09157}.

\bibitem[{Zhao et~al.(2020)Zhao, Vemulapalli, Mansfield, Gong, Green, Shapira,
  and Wu}]{zhao2020contrastive}
Zhao, X.; Vemulapalli, R.; Mansfield, P.; Gong, B.; Green, B.; Shapira, L.; and
  Wu, Y. 2020.
\newblock Contrastive Learning for Label-Efficient Semantic Segmentation.
\newblock \emph{arXiv preprint arXiv:2012.06985}.

\bibitem[{Zhou et~al.(2021)Zhou, Sodha, Pang, Gotway, and
  Liang}]{zhou2021models}
Zhou, Z.; Sodha, V.; Pang, J.; Gotway, M.~B.; and Liang, J. 2021.
\newblock Models Genesis.
\newblock \emph{Medical Image Analysis}, 67: 101840.

\bibitem[{Zhou et~al.(2019)Zhou, Sodha, Rahman~Siddiquee, Feng, Tajbakhsh,
  Gotway, and Liang}]{zhou2019models}
Zhou, Z.; Sodha, V.; Rahman~Siddiquee, M.~M.; Feng, R.; Tajbakhsh, N.; Gotway,
  M.~B.; and Liang, J. 2019.
\newblock Models Genesis: Generic Autodidactic Models for 3D Medical Image
  Analysis.
\newblock 384--393. Cham: Springer International Publishing.
\newblock ISBN 978-3-030-32251-9.

\bibitem[{Zhu et~al.(2019)Zhu, Zhang, Qiu, Liu, Liu, and
  Chen}]{zhu2019comparison}
Zhu, J.; Zhang, J.; Qiu, B.; Liu, Y.; Liu, X.; and Chen, L. 2019.
\newblock Comparison of the automatic segmentation of multiple organs at risk
  in CT images of lung cancer between deep convolutional neural network-based
  and atlas-based techniques.
\newblock \emph{Acta Oncologica}, 58(2): 257--264.

\bibitem[{Zhuang et~al.(2019)Zhuang, Li, Hu, Ma, Yang, and
  Zheng}]{zhuang2019self}
Zhuang, X.; Li, Y.; Hu, Y.; Ma, K.; Yang, Y.; and Zheng, Y. 2019.
\newblock Self-supervised feature learning for 3d medical images by playing a
  rubik’s cube.
\newblock In \emph{International Conference on Medical Image Computing and
  Computer-Assisted Intervention}, 420--428. Springer.

\end{thebibliography}
\end{document}